\documentclass[fp, twocolumn]{jpsj3}

\allowdisplaybreaks[1]

\usepackage{txfonts}
\usepackage{graphicx}
\usepackage{bm}
\usepackage{braket}
\usepackage{cases}
\usepackage{mathrsfs}
\usepackage{amsmath}

\usepackage{color}
\usepackage{soul}

\usepackage{float}

\title{Magnetic-Field-Driven Antiferromagnetic Domain Wall Motion}
\date{\today}
\author{Jotaro J.~Nakane and Hiroshi Kohno}
\inst{Department of Physics, Nagoya University, Nagoya 464-8602, Japan} 

\abst{
  We theoretically study the antiferromagnetic domain wall motion actuated 
by an inhomogeneous external magnetic field. 
  The Lagrangian and the equations of motion of antiferromagnetic spins  
under an inhomogeneous magnetic field are derived,
first in terms of the N\'eel vector, 
and then using collective coordinates of the domain wall. 
 A solution is found that describes the actuation of a domain wall by an inhomogeneous field, 
in which the motion is initiated by a paramagnetic response of wall magnetization,  
which is then driven by a Stern-Gerlach like force. 
  The effects of pinning potential are also investigated. 
  These results are in good agreement with atomistic simulations. 
 While the present formulation contains the so-called intrinsic magnetization 
associated with N\'eel texture, 
a supplementary discussion is given to reformulate the theory in terms of physical magnetization   
without the intrinsic magnetization. 
}

\begin{document}

\maketitle

\section{Introduction}
 Antiferromagnets have recently gained high expectations as a candidate for next generation spintronic devices,
 owing to its robustness to external magnetic field, THz range spin dynamics, and the variety of hosting materials \cite{Jungwirth2016,Qiu2016,Macdonald2011,sinova2017,Baltz2018}. 
 The demonstration of antiferromagnetic domain switching using the N\'eel spin-orbit torque \cite{wadley2016} encouraged researchers to pursue antiferromagnetic spintronics. 
 Still, despite its appealing features and high publicity, 
 the manipulation and detection of spin textures in antiferromagnets remain a challenge 
due to its unsusceptible nature. 
 
 A domain wall is one of the topological objects in antiferromagnets 
that may prove useful in memory devices,
 and its creation, manipulation and detection has been the scope in numerous studies.
 Some pioneering theoretical works show that antiferromagnetic domain walls can be driven 
by spin waves \cite{Tveten2014} and spin-orbit torques \cite{Gomonay2016,Shiino2016},
 though experimental reports on antiferromagnetic domain wall motion are currently limited 
to indirect observations \cite{Herranz2009}. 
 Domain walls in materials similar to antiferromagnets, such as synthetic antiferromagnets\cite{Yang2015} 
and ferrimagnets around the angular momentum compensation temperature 
\cite{Kim2017,Okuno2019}, have also been studied, 
which allow for an easier observation and manipulation of domain walls.

 Recently, it was proposed that antiferromagnetic spin textures give rise to intrinsic magnetization \cite{Tveten2016}.
 It was demonstrated that this intrinsic magnetization couples to external magnetic fields, 
 and may be used to actuate antiferromagnetic domain wall motion.  
 In this paper, we reinvestigate this problem starting from 
  the same model. 
 We found an additional coupling of the N\'eel vector to the inhomogeneous magnetic field, 
similar to the one in Ref.~\cite{Yuan2018}, 
which nullifies the effect of the above intrinsic magnetization. 
  With the new Lagrangian obtained,  
we find an alternative mechanism for domain wall motion actuated 
by an inhomogeneous external magnetic field.

 This paper is organized as follows. 
 After presenting in Sec.~2 the Lagrangian and equations of motion for the antiferromagnetic order parameter
 (N\'eel vector) under an inhomogeneous magnetic field, 
we derive in Sec.~3 the equations of motion in terms of collective coordinates of a domain wall.
 By solving the equations, we find 
a solution in which the domain wall position grows exponentially with time. 
 Interestingly, there is no domain wall motion in the absence of damping.
 We also study the effects of pinning introduced by a local modulation of easy-axis magnetic anisotropy. 
 Finally, we perform an atomistic simulation to test the analytical results, 
and see that they are in good agreement.
 As a supplementary discussion, we identify the physical magnetization and reformulate the theory
therewith.

\section{Model}

 In this section, we derive an effective Lagrangian that describes low-frequency, 
long-wavelength spin dynamics of an antiferromagnet, starting from a lattice spin model.  
 We closely follow the procedure described in Ref.~\cite{Tveten2016}, 
except that we consider the inhomogeneity of the external magnetic field 
from the beginning of the formulation.

\subsection{Hamiltonian}

 We start with a Heisenberg Hamiltonian for classical spins on a one-dimensional lattice 
with antiferromagnetic exchange coupling $J>0$, easy-axis anisotropy $K>0$, 
and Zeeman coupling, 
\begin{align}
  H &= J\sum_i {\bm S}_{i}\cdot {\bm S}_{i+1}
  -K\sum_i ({\bm S}_i\cdot \bm e_z)^2
  + \gamma \hbar \sum_i {\bm H}_i \cdot {\bm S}_i  ,
\end{align}
where $\gamma$ is the gyromagnetic ratio. 
 The localized spins and magnetic field at lattice site $i$ are written as ${\bm S}_i$ and ${\bm H}_i$, respectively.
 In a typical easy-axis antiferromagnet, the exchange energy dominates $J \gg K$, giving rise to 
relatively thick domain walls (e.g. $150{\rm nm}$ for NiO \cite{Weber2003}). 
 Therefore, we work in the exchange approximation $J \gg K$ (and $J \gg \gamma \hbar |H_i|$), 
and focus on spin textures with slow spatial/temporal variations.

 Let us write the antiferromagnetic spins in terms of the N\'eel and uniform moments, 
\begin{align}
  {\bm n}_n &= \frac{{\bm S}_{2n}-{\bm S}_{2n+1}}{2S} , 
\ \ \ \ \ 
  {\bm l}_n  = \frac{{\bm S}_{2n}+{\bm S}_{2n+1}}{2S} , 
\label{eq:nl} 
\end{align}
respectively, where $|{\bm S}_{i}|=S$ is a constant. ($i$ is the site index, and $n$ is the unit-cell index.)
The original spins are retrieved by 
\begin{align}
  {\bm S}_{2n} &= S({\bm l}_n+{\bm n}_n) , \ \ \  
  {\bm S}_{2n+1} = S({\bm l}_n-{\bm n}_n) . 
\label{eq:S}
\end{align}
 With ${\bm n}_n$ and ${\bm l}_n$, the Hamiltonian is written as  
$H = \sum_n h_n$, where 
\begin{align}
 h_n &=
  JS^2   \Biggl\{ 
  2 ({\bm l}_n^2 - {\bm n}_n^2 ) 
  - \frac{({\bm l}_n - {\bm l}_{n-1})^2}{2}
  + \frac{({\bm n}_n - {\bm n}_{n-1})^2}{2}
\nonumber \\
&\qquad\qquad\quad
  + ({\bm n}_n - {\bm n}_{n-1}) \cdot {\bm l}_n
  - {\bm n}_n \cdot ({\bm l}_n - {\bm l}_{n-1})  \Biggr\} 
\nonumber \\
&\quad
  -2KS^2  \bigl\{  (l_n^z)^2 + (n_n^z)^2 \bigr\} 
\nonumber \\
&\quad
  + \gamma \hbar S \Bigl\{ ({\bm H}_{2n} + {\bm H}_{2n+1})\cdot{\bm l}_n 
                  - ({\bm H}_{2n+1} - {\bm H}_{2n}) \cdot {\bm n}_n \Bigr\} . 
\end{align}
 We adopt the continuum approximation and write 
 ${\bm n}_n - {\bm n}_{n-1} \simeq 2a \partial_x{\bm n}$ and 
 ${\bm l}_n - {\bm l}_{n-1} \simeq 2a \partial_x{\bm l}$, where $a$ is the lattice constant. 
 The magnetic field is also assumed slowly-varying (having no staggered component) 
and thus  
 ${\bm H}_{2n} + {\bm H}_{2n+1} \simeq 2{\bm H}$,
 ${\bm H}_{2n+1} - {\bm H}_{2n} \simeq a \partial_x{\bm H}$,
 and the summation is replaced by an integration
 ${\displaystyle\sum_n = \int \frac{dx}{2a}}$. 
 Thus the Hamiltonian is written as
\begin{align}
  H
&= JS^2\int \frac{dx}{2a} \bigg\{ 
  2({\bm l}^2 - {\bm n}^2) 
  +\frac{(2a)^2}{2} \bigl[ (\partial_x{\bm n})^2 - (\partial_x{\bm l})^2 \bigr] 
\nonumber \\
&\qquad\qquad\quad
  +(2a) \bigl[ {\bm l} \cdot (\partial_x{\bm n})
  - {\bm n}\cdot (\partial_x{\bm l}) \bigr] 
  \bigg\}
\nonumber \\
&\quad
  - 2KS^2\int \frac{dx}{2a} \, \big\{ (l^z)^2 +(n^z)^2 \big\}
\nonumber \\&\quad
  + \gamma \hbar S 
  \int \frac{dx}{2a} \big\{ 2{\bm H} \cdot {\bm l} - a (\partial_x{\bm H}) \cdot{\bm n} \big\} . 
\label{eq:cont_hamiltonian}
\end{align}
 As we shall see later, $|{\bm l}|=O(a/\lambda)$ in the exchange approximation ($J \gg K$), 
 where $\lambda = a \sqrt{J/2K}$ is the typical length scale of spatial variation.
 We discard the terms which are of higher order in $l$, such as $K l^2=O(J l^4)$  
and $(a \, \partial_x \bm l)^2=O(l^4)$.
 Using the constraints, 
\begin{align}
  {\bm n}^2 + {\bm l}^2 &= 1  , \ \ \ 
  {\bm n} \cdot {\bm l}  = 0  , 
\label{eq:constraints}
\end{align}
or ${\bm n}\cdot\partial_x{\bm l} = - (\partial_x {\bm n})\cdot{\bm l}$, 
which follow from $|{\bm S}_{i}|= {\rm const.}$, we obtain
\begin{align}
  H
&\simeq
  4JS^2 \int \frac{dx}{2a} \,
  \bigg\{   {\bm l}^2 + \frac{a^2}{2} (\partial_x{\bm n})^2
    + a \, {\bm l} \cdot (\partial_x {\bm n})  - \frac{K}{2J} (n^z)^2
  \bigg\}
\nonumber \\&\quad
  + \gamma \hbar S 
    \int \frac{dx}{2a} \left( 2{\bm H} \cdot {\bm l} - a(\partial_x{\bm H}) \cdot {\bm n} \right)   
.
\label{eq:H}
\end{align}
 As seen, the magnetic field couples not only to ${\bm l}$ but also to 
${\bm n}$.\cite{Yuan2018}

\subsection{Lagrangian and Equations of Motion}

 To derive the Lagrangian, $L = L_0 - H$, we next look at its kinetic part, 
\begin{align}
   L_0 
&= \hbar S \sum_i  \dot\phi_i\cos\theta_i
\nonumber \\
&= \hbar S\sum_n  \left(\dot\phi_{2n}\cos\theta_{2n}+ \dot\phi_{2n+1}\cos\theta_{2n+1}\right)  , 
\end{align}
where the spins are expressed as
\begin{equation}
  {\bm S}_i =  S( \sin\theta_i \cos\phi_i , \sin\theta_i \sin\phi_i , \cos\theta_i ) .
\end{equation}
 Let 
$\theta_{2n+1} = \pi - (\theta_{2n}+\delta\theta_{2n+1})$
and
$\phi_{2n+1} = \pi + (\phi_{2n}+\delta\phi_{2n+1})$,
so that the neighboring spins are totally antiparallel when $\delta\theta=\delta\phi=0$.
 To leading order in $\delta \theta$ and $\delta\phi$, one finds\cite{Dombre1988,Haldane1988}  
\begin{equation}
   L_0  =   2\hbar S \int \frac{dx}{2a} \, {\bm l} \cdot ( {\bm n} \times \dot {\bm n})  , 
\label{eq:L_0}
\end{equation}
up to a total time derivative. 
 This shows that $2\hbar S ({\bm l} \times {\bm n})$ is the canonical momentum 
conjugate to ${\bm n}$. 
 As a side note, the emergent gauge field, 
$A_{{\rm AF}, i} = {\bm l}\cdot(\partial_i{\bm n}\times {\bm n})$, demonstrated in \cite{Nakane2019} 
for canted antiferromagnets 
complies with this kinetic term.

 Damping is taken into account by Rayleigh's dissipation function, 
\begin{align}
  W = \alpha \frac{\hbar}{2S}   \sum_{i}\dot{{\bm S}}_i^2 
=  \alpha \hbar S \sum_{n} ( \dot{\bm l}_n^2+\dot{\bm n}_n^2 ) . 
\label{eq:rayleigh_dissipation}
\end{align}
 In the continuum approximation, we write 
\begin{align}
  W &= 2 \hbar S \int \frac{dx}{2a} 
 \left(  \frac12 \alpha_l \dot{\bm l}^2 + \frac12 \alpha_n \dot{\bm n}^2  \right) , 
\label{eq:rayleigh_dissipation_neel}
\end{align}
where we introduced two damping constants, $\alpha_l$ and $\alpha_n$, 
for more generality \cite{Hals2011}.

\begin{figure}[tbp]
  \centering
  \includegraphics[width=0.48\textwidth]{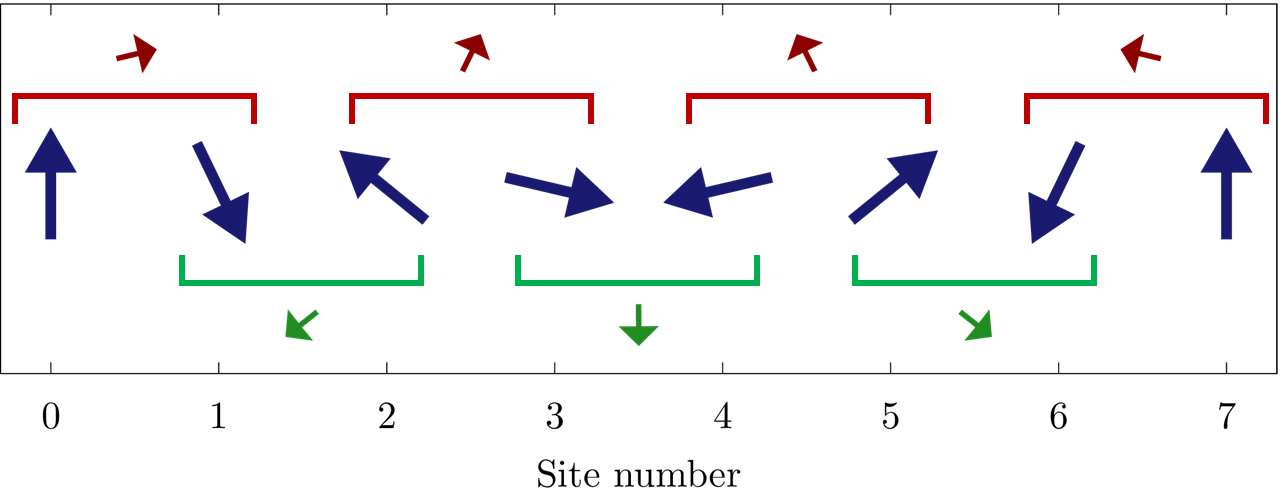}
  \caption{(Color online) 
    This figure illustrates that the uniform moment, 
     Eq.~(\ref{eq:nl}),  in the presence of 
     antiferromagnetic spin texture depends on the choice of unit cell. 
    The long arrows represent the atomic spins, ${\bm S}_i$. 
    The top red arrows and the bottom green arrows represent 
     the \lq\lq uniform moments'' locally defined by the average of the two 
    neighboring spins, $({\bm S}_i + {\bm S}_{i+1})/2$. 
    Here, they arise from the antiferromagnetic spin texture, and as seen, they are mutually opposite. 
    Such unit-cell-choice dependent components should not represent the physical magnetization. 
  }
\label{fig:afm_fig1}
\end{figure}

 By noting the constraints,  Eq.~(\ref{eq:constraints}), the equations of motion are obtained as 
\begin{align}
  \left\{
  \begin{array}{cc}
    \dot{\bm n}
&= \displaystyle
     \left( \frac{1}{s_0} \frac{\delta H}{\delta {\bm n}} + \alpha_n \dot{\bm n} \right) \times {\bm l}
    + \left( \frac{1}{s_0} \frac{\delta H }{\delta{\bm l}} + \alpha_l \dot{\bm l} \right) \times {\bm n}
\\[8pt]
  \dot{\bm l}
&= \displaystyle
     \left( \frac{1}{s_0} \frac{\delta H}{\delta {\bm n}} + \alpha_n \dot{\bm n} \right) \times {\bm n}
    + \left( \frac{1}{s_0} \frac{\delta H }{\delta{\bm l}} +\alpha_l \dot{\bm l} \right) \times {\bm l}
  \end{array}
  \right. , 
\label{eq:afm_eom_sym}
\end{align}
in agreement with Ref.~\cite{Brataas2015}. 
 Here, we defined the angular momentum density $s_0 = 2 \hbar S/(2a)$. 
 Note that these equations of motion respect the constraints, Eq.~(\ref{eq:constraints}). 
 The first equation of  Eq.~(\ref{eq:afm_eom_sym}) can be written as \cite{Tveten2016,Ivanov1995}
\begin{align}
  {\bm l} 
&=   \frac{\hbar }{4JS}  ({\bm n} \times \dot {\bm n} - \gamma {\bm H}_\perp)
   - \frac{a}{2} (\partial_x {\bm n})  , 
\label{eq:afm_eom_uniform}
\end{align}
to leading order in ${\bm l}$, 
where ${\bm H}_\perp = {\bm n} \times ({\bm H} \times {\bm n})$ is the component 
perpendicular to ${\bm n}$. 
 The first term ($\sim {\bm n} \times \dot {\bm n}$) embodies the momentum nature of ${\bm l}$ 
conjugate to ${\bm n}$, 
 namely, it is proportional to the \lq\lq velocity'' $\dot {\bm n}$. 
 The second term is due to canting induced by ${\bm H}$. 
 The linear dependence on the field ${\bm H}_\perp$ 
indicates that the response to it is paramagnetic. 
The last term  
($\sim \partial_x {\bm n}$) is referred to in Ref.~\cite{Tveten2016}  
as the intrinsic magnetic moment induced by the N\'eel texture.
 Substituting this result into the second equation of Eq.~(\ref{eq:afm_eom_sym}), 
one obtains the equation of motion written solely by the N\'eel vector, 
\begin{align}
  -\frac{\hbar^2  }{2J} \, \ddot{\bm n}\times {\bm n}
&= \biggl[
    - 2J S^2 a^2 (\partial^2_x {\bm n}) 
    - 4KS^2 ({\bm n} \cdot \hat z) \, \hat z
    + 2 \hbar S \alpha_n \dot{\bm n}
\nonumber\\
&\quad 
    + \frac{\gamma \hbar^2 }{J}({\bm H} \cdot {\bm n}) \, \dot{\bm n} \times {\bm n}
    +  \frac{(\gamma \hbar)^2}{2J}    ({\bm H} \cdot {\bm n}) \, {\bm H}
\nonumber\\
&\quad 
    - \frac{ \gamma \hbar^2}{2J}  (\dot{\bm H} \times {\bm n})
  \biggr] \times {\bm n}  . 
\label{eq:afm_eom_neel}
\end{align}
 As an important observation, the terms with $\partial_x{\bm H}$ have been canceled out.   
 To see what happened, let us go back to the Lagrangian $L$, or its density, 
\begin{align}
  {\cal L}  &=  s_0 \, {\bm l} \cdot ( {\bm n} \times \dot {\bm n}) 
 - s_0 \gamma 
   \left\{  {\bm H} \cdot {\bm l} -  \frac{a}{2} (\partial_x{\bm H}) \cdot {\bm n} \right\} 
\nonumber \\
& - \frac{2JS^2}{a}  
  \bigg\{   {\bm l}^2 + \frac{a^2}{2} (\partial_x{\bm n})^2
    + a \, {\bm l} \cdot (\partial_x {\bm n})  - \frac{K}{2J} (n^z)^2
  \bigg\} . 
\label{eq:L_density}
\end{align}
 Since this is quadratic in ${\bm l}$, one can \lq\lq integrate out'' ${\bm l}$ 
and obtain the one written solely by the N\'eel vector, 
\begin{align}
 {\cal L}  
&= \frac{\hbar^2 }{8Ja} \dot{\bm n}^2
  - \frac{JS^2 a}{2}   (\partial_x \bm n)^2
  + \frac{KS^2}{a} (n^z)^2
\nonumber \\
&\qquad
  + \frac{(\gamma \hbar)^2}{8Ja}({\bm H} \times \bm n)^2 
  + \frac{\gamma \hbar S}{2}   \, \partial_x ({\bm H} \cdot {\bm n})  
\nonumber \\
&\qquad
  \, - \, \frac{s_0}{2}  ({\bm n} \times \dot {\bm n}) \cdot \left[ a \, \partial_x {\bm n} 
                           + \frac{\gamma \hbar}{2JS} {\bm H} \right]  . 
\label{eq:Lagrangian}
\end{align}
 Here, the Zeeman coupling of the \lq\lq intrinsic magnetization''  
($\sim {\bm H} \!\cdot\! \partial_x {\bm n}$) has been combined with the additional term\cite{Yuan2018}  
($\sim {\bm n} \!\cdot\! \partial_x {\bm H}$ in Eq.~\eqref{eq:cont_hamiltonian}), 
forming a total derivative, $\partial_x ({\bm H} \cdot {\bm n})$. 
 This is why the intrinsic moment does not appear in the equation of motion, 
Eq.~(\ref{eq:afm_eom_neel}). 
 Intuitively, this can be understood from Fig.~\ref{fig:afm_fig1}, 
which shows that the texture-induced uniform moment depends on the unit-cell choice; 
if another choice is made, it changes sign. 
 This means that the texture-induced uniform moment is an artifact of the 
parametrization of Eq.~(\ref{eq:nl}), and does not appear in physical phenomena.

 Two small notes. 
 First, the quadratic term in ${\bm H}$, namely,  
the fifth term $\sim ({\bm H} \cdot {\bm n}) \, {\bm H}$ in Eq.~(\ref{eq:afm_eom_neel}) 
or the fourth term $\sim ({\bm H} \times \bm n)^2$ in Eq.~(\ref{eq:Lagrangian}), 
has the same form as the magnetic anisotropy term, 
hence the magnetic field acts as a hard-axis anisotropy. 
(But the effect is small, see below.) 
 Second, the sixth term $\sim ({\bm n} \times \dot {\bm n} ) \cdot \partial_x {\bm n} $ 
in Eq.~(\ref{eq:Lagrangian}) is \lq\lq topological'',  
and does not contribute to the equation of motion, hence can be omitted.

\section{Domain Wall Motion}

 In this section, we study the domain wall motion in an inhomogeneous magnetic field 
using collective coordinates. 
 We take the magnetic field to be in the easy-axis ($\hat z$-) direction 
with magnitude linearly varying in space, 
\begin{equation}
  {\bm H} = H_z \hat z = (H_0 + H_1 x) \, \hat z . 
\label{eq:H(x)}
\end{equation}

\subsection{Static Domain Wall Solution}
To work with collective coordinates of the antiferromagnetic domain wall, 
one must first obtain a static domain wall solution.
 Dropping the time derivative terms, Eq.~(\ref{eq:afm_eom_neel}) becomes 
\begin{align}
  0 &=  \left[  - 2J   a^2 (\partial^2_x {\bm n})
    - 4K ({\bm n}\cdot\hat z) \, \hat z
    + \frac{(\gamma \hbar)^2}{2J S^2} ({\bm H} \cdot {\bm n}) \, {\bm H}
  \, \right] \times {\bm n}  .
\end{align}
 With Eq.~(\ref{eq:H(x)}) for the magnetic field, 
\begin{align}
  0 &=  \left[  - 2J   a^2 (\partial^2_x {\bm n}) 
   - 4K' ({\bm n} \cdot \hat z) \, \hat z \, \right] \times {\bm n}  ,
\end{align}
where
$ K' = K- (\gamma \hbar H_z)^2 /(8JS^2)$.
 Note that the Zeeman coupling $\gamma \hbar H_z$ is typically of the order of few kelvins 
(for $H_z$ of few tesla), similar to the anisotropy energy. 
 Thus, the difference between $K'$ and $K$ is of order $O(K^2/J)$ 
which is dismissed under our current approximation, $ K'  \simeq K$. 
 We write the N\'eel vector using the polar and azimuthal angles, 
\begin{equation}
  {\bm n} = ( \sin\theta \cos\phi , \sin\theta \sin\phi , \cos\theta ) ,
\end{equation}
under the assumption that the uniform moment is small 
(namely, $1 = {\bm n}^2 + {\bm l}^2 \simeq {\bm n}^2$). 
 Assuming $\phi$ is spatially uniform,  and noting that 
$\hat z = {\bm n} \cos\theta  - {\bm e}_\theta \sin\theta $,
the static texture satisfies
\begin{align}
  0  &=  -\lambda^2\partial_x^2\theta  + \cos\theta\sin\theta  , 
\end{align}
where $\lambda = a \sqrt{J/2K}$.  
 A domain wall solution is given by \cite{Tatara2008}
\begin{align}
  \cos\theta &= \tanh\left( \pm \frac{x-X}{\lambda} \right) ,   
\label{eq:walker_dw}
\end{align}
and thus $\sin\theta = \left[ \cosh\frac{x-X}{\lambda} \right]^{-1}$, 
where $X$ is the domain wall position. 
The $\pm$ sign is the topological charge of the domain wall. 
 The position of the wall, $X$, and the angle of the wall plane, $\phi$ ($=$ const.),  
will be promoted to dynamical variables in the next subsection.

\subsection{Collective Coordinate Description}

 Using the domain wall solution for ${\bm n}$, Eq.~(\ref{eq:walker_dw}),  
the kinetic part of the Lagrangian, $L_0$ in Eq.~(\ref{eq:L_0}), is obtained as
\begin{align}
 L_{{\rm DW},0} &= 2 s_0 \bigl( \pm l_\phi \dot X - \lambda l_\theta \dot \phi  \bigr) ,
\label{eq:L_DW}
\end{align} 
where we defined 
\begin{align}
  l_\theta  
&=  \int \frac{dx}{2\lambda}  \frac{ {\bm e}_\theta \cdot {\bm l}(x)}{\cosh \frac{x-X}{\lambda}}  , 
\ \ \ 
   l_\phi  
 =  \int \frac{dx}{2\lambda}  \frac{ {\bm e}_\phi \cdot {\bm l}(x)}{\cosh \frac{x-X}{\lambda}} . 
\end{align}
 Equation (\ref{eq:L_DW}) indicates that $l_\theta$ and $l_\phi$ 
are canonical momenta conjugate to $\phi$ and $X$, respectively,  
which should be considered as new collective variables. 
 Note that a uniform moment induced by the (longitudinal) field,  Eq.~(\ref{eq:H(x)}), 
is localized at the domain wall (see Eq.~(\ref{eq:afm_eom_uniform})). 
 In a more systematic treatment, this corresponds to expanding $\bm l$ with some complete set 
of functions $\{ \varphi_n (x) \}$ \cite{Tatara2008}, 
\begin{align}
 {\bm l}(x)
 &=  (l_\theta {\bm e}_{\theta}+l_\phi{\bm e}_{\phi}) \, \varphi_0 (x) + \sum_k {\bm l}_k \, \varphi_k (x) , 
\end{align}
where
\begin{align}
  \varphi_0 (x) =   \frac{1}{\cosh\frac{x-X}{\lambda}} ,
\end{align}
and retain the first two terms. 
 See Ref.~\cite{Tatara2008} for other $\varphi_k (x)$'s, 
which, together with $\varphi_0 (x)$, form a complete orthogonal basis.

 The Hamiltonian, Eq.~(\ref{eq:H}), then becomes
\begin{align}
  H_{\rm DW} &= 4JS^2\frac{\lambda}{a} 
    \left\{   \left( l_\theta \mp \frac{a}{2\lambda} \right)^2 + l_\phi^2 
            +  \frac{a^2}{2\lambda^2}  \right\}  
\nonumber\\
&\qquad
  - \gamma \hbar S \biggl( \frac{2\lambda}{a} \, l_\theta  \mp 1 \biggr)  (H_0 + H_1 X) . 
\label{eq:H_DW}
\end{align}

 The dissipation function Eq.~(\ref{eq:rayleigh_dissipation_neel}), 
dismissing the $\alpha_l \, \dot{\bm l}^{\, 2}$ term \cite{Tveten2013}, is given by
\begin{align}
  W_{\rm DW} &= s_0 \alpha_n \lambda \, 
  \biggl( \frac{\dot{X}^2}{\lambda^2}  + \dot{\phi}^2 \biggr) .
\end{align}

 These results lead to the following four equations of motion, 
\begin{align}
 \pm \dot{l}_\phi &=  - \alpha_n \dot{X} / \lambda
          + \gamma H_1 \lambda \left(  l_\theta \mp \frac{a}{2\lambda} \right)  , 
\label{eq:eom_X}
\\
   \dot{l}_\theta  &=    \alpha_n \dot{\phi}  , 
\label{eq:eom_phi}
\\
   \dot \phi  &=   
   - (4J S /\hbar) \left( l_\theta  \mp \frac{a}{2\lambda} \right) 
   + \gamma (H_0 + H_1 X) , 
\label{eq:eom_ell_phi}
\\
     \dot X &= \pm (4J S /\hbar)  \lambda l_\phi . 
\label{eq:eom_ell_theta}
\end{align} 
Note that $(X, l_\phi )$ and $(\phi, l_\theta)$ are coupled via $H_1$. 
 However, 
$l_\theta$ and $l_\phi$ can be eliminated, resulting in two coupled equations 
for $X$ and $\phi$, or $\chi$ and $\varphi$ defined by 
\begin{align}
  \chi &= \frac{X}{\lambda}+ \frac{H_0}{H_1\lambda} , 
\ \ \ \ \  
  \varphi  = \phi  + \phi_1 , 
\end{align}
as
\begin{align}
  \left\{
  \begin{array}{ll}
    \ddot{\chi}  &=  - \tilde \alpha \dot{\chi}+ \tilde H_1 \tilde \alpha \varphi
  \\
    \dot{\varphi} &=  \tilde H_1\chi  - \tilde \alpha \varphi
  \end{array}
  \right.  . 
\end{align} 
 Here, we defined
\begin{align}
  \tilde \alpha &= \frac{4 S  J}{\hbar} \alpha_n , 
\ \ \ \ \ 
  \tilde H_1  = \gamma H_1 \lambda , 
\end{align}
and 
$\phi_1= \alpha_n^{-1} ( l_{\theta}^0 \mp a/2\lambda) - \phi_0 $, 
where $l_{\theta}^0$ and $\phi_0$ are initial values introduced 
when Eq.~(\ref{eq:eom_phi}) is integrated in time. 
 It is readily seen that the acceleration of the domain wall is absent if there is no damping, 
$\tilde \alpha = 0$. 
 This feature is not seen in Refs.~\cite{Tveten2016,Yuan2018}.
 Writing the equations of motion in matrix form, 
\begin{align}
  \frac{d}{dt}
  \begin{pmatrix}
    \dot{\chi} \\
    \chi \\
    \varphi
  \end{pmatrix}
&=
  \begin{pmatrix}
    - \tilde \alpha & 0 & \tilde H_1 \tilde \alpha \\
    1 & 0 & 0 \\
    0 & \tilde H_1 & - \tilde \alpha
  \end{pmatrix}
  \begin{pmatrix}
    \dot{\chi} \\
    \chi \\
    \varphi
  \end{pmatrix}  , 
\end{align}
and assuming the solution of the form $\sim e^{\varepsilon t}$, 
the problem reduces to an eigenvalue problem with determinant, 
\begin{align}
  \varepsilon (\varepsilon + \tilde \alpha )^2  - \tilde \alpha  \tilde H_1^2  = 0 .
\end{align}
 Since $\tilde \alpha$ is positive, this equation has one real positive root $\varepsilon_0$, 
and two complex roots $\varepsilon_1$ and $\varepsilon_2$ ($= \varepsilon_1^*$) 
with negative real parts.
 Using Cardano's method, the roots are written as
\begin{align}
  \varepsilon_n 
&=  -\frac{2\tilde \alpha}{3}
  + \omega^n \, \sqrt[3]{  q + \sqrt{q^2-p^3 } }
  + \omega^{2n}\, \sqrt[3]{ q - \sqrt{q^2-p^3 } }  ,
\nonumber\\
&\qquad\qquad 
  n=0,\,1,\,2
\label{eq:varepsilon}
\end{align}
where
$p = \frac{1}{9} \tilde \alpha^2$, 
$q = ( \frac{1}{27} {\tilde \alpha}^2 + \frac{1}{2} \tilde H_1^2 ) \tilde \alpha$, 
$\omega = -\frac12 + \frac12\sqrt{3} i$, 
and the real branch of the cube roots are chosen. (Note that $q^2-p^3 \geq 0$.)
 With these roots, the general solution is given by 
\begin{align}
  \chi
&=  C_0 e^{\varepsilon_0 t} + {\rm Re} \left[ C_1e^{\varepsilon_1 t} \right] , 
\label{eq:dw_particle_function}
\\
  \varphi 
&=  C_0 \frac{\tilde H_1 }{\tilde \alpha + \varepsilon_0} e^{\varepsilon_0 t} 
   + {\rm Re} \left[ C_1\frac{\tilde H_1 }{\tilde \alpha + \varepsilon_1} e^{\varepsilon_1 t} \right] , 
\end{align}
where a real constant $C_0$ and a complex constant $C_1$ are determined by initial conditions. 
 Thus, we find that an inhomogeneous magnetic field drives domain wall motion, 
$X \sim \lambda C_0 e^{\varepsilon_0 t}$, that grows exponentially in time.

\begin{figure}[t]
  \centering
  \includegraphics[width=0.48\textwidth]{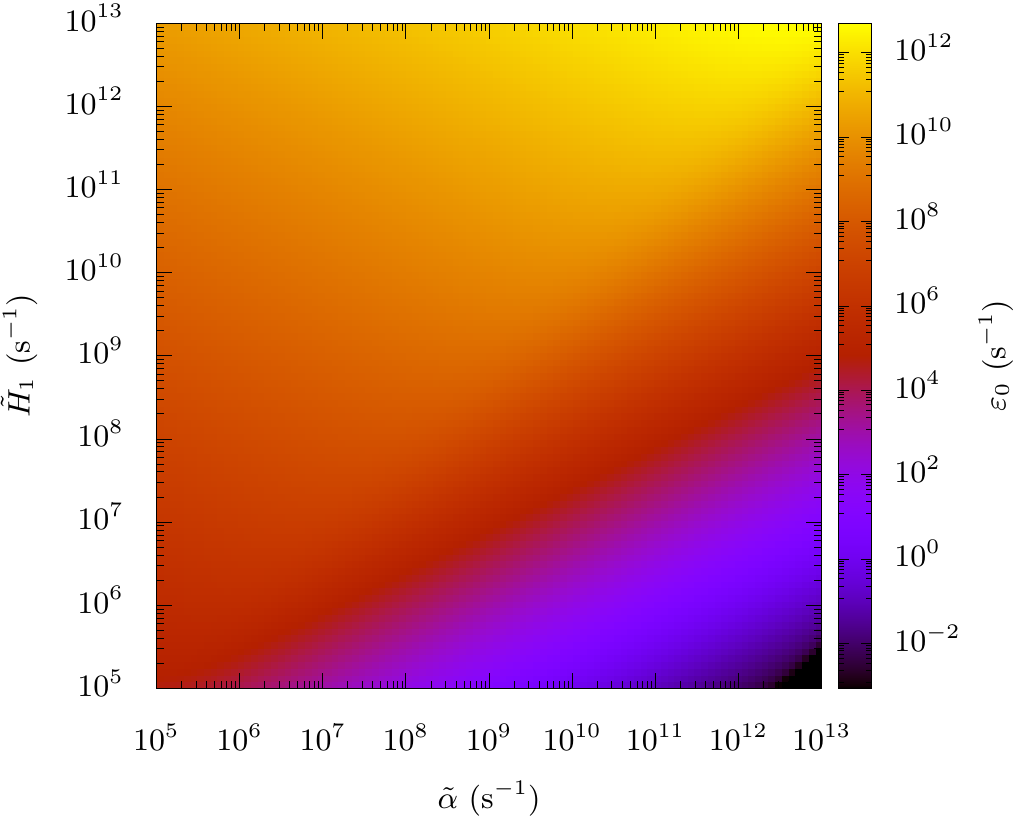}
  \caption{(Color online) 
    The real positive eigenvalue $\varepsilon_0$ [Eq.~(\ref{eq:varepsilon})]  
    plotted as a function of $\tilde \alpha$ and $\tilde H_1$. 
    It increases with $\tilde H_1$ as expected.
  }
\label{fig:afm_fig2}
\end{figure}

When $\tilde \alpha \gg\tilde H_{1}$, the roots can be given as 
\begin{align}
  \varepsilon_0 &\simeq  \frac{3\tilde H_1^2}{\tilde \alpha} , \ \ \ \ \ 
  \varepsilon_1 \simeq  -\tilde \alpha + i\tilde H_1 , 
\end{align}
to leading order of $\tilde \alpha/\tilde H_1$.
 We expect most antiferromagnets under magnetic field satisfy this condition. 
For example, for $2S \alpha_n = 10^{-3}$, $J = 10^3 {\rm K}$, 
$\gamma \hbar H_1 = 1 {\rm K}/{\rm cm}$, $\lambda = 100{\rm nm}$, and $S = 1$,
then, $\tilde \alpha = 10^{11}{\rm s}^{-1}$, $\tilde H_1 = 10^{5}{\rm s}^{-1}$, and $\varepsilon_0 = 1 {\rm s}^{-1}$.
A plot of $\varepsilon_0$ in  Eq.~(\ref{eq:varepsilon}) is shown in Fig.~\ref{fig:afm_fig2} 
as a function of $\tilde \alpha$ and $\tilde H_1$.

 The above solutions (and equations) are limited to the case, 
$| l_\theta | , | l_\phi | \ll 1$, in order for the exchange approximation to be valid. 
 Since $l_\theta / \chi \sim \alpha_n \tilde H_1 /(\tilde \alpha + \varepsilon_0)$, 
this requires $|\chi | \ll (\tilde \alpha + \varepsilon_0) / (\alpha_n |\tilde H_1|)$, or 
$| X - X_0| / \lambda \ll {\rm max}\{ 4SJ /\hbar, \varepsilon_0/\alpha_n \} / |\tilde  H_1|  \equiv b$, 
namely, the domain wall position should be within the distance $\sim \frac{1}{10} b \lambda$ 
from the position $X_0 \equiv - H_0/H_1$ of vanishing external field, $H_z = 0$. 
 Beyond this point, the nonlinearity of $l_\theta$ may not be neglected. 
 With parameters given above, we have $b \sim 10^9$, 
hence $b \lambda$ is much larger than the domain wall width,
and the above condition is always satisfied. 
 We note that the above condition is equivalent to $\gamma \hbar |H_z| \ll J$ 
at the domain wall position, which is practically always satisfied.

\subsection{With Pinning Potential}

 We introduce a pinning potential by a local modulation $\delta K$ of the anisotropy $K$ 
at the origin, $K \to K - 2a \delta (x) \delta K $ \cite{Tatara2008}. 
 The pinning potential of the domain wall is then given by 
\begin{align}
  V_{\rm pin} 
&= - 2 \delta K S^2 \frac{1}{\cosh^2 ( X/\lambda )}
\nonumber \\
&\simeq  2 \delta K S^2 ((X/\lambda)^2 - 1) \, \Theta (\lambda-|X|)  , 
\label{eq:Vpin}
\end{align}
which we approximated by a truncated parabola. 
 ($\Theta$ is the Heaviside step function.)
 When $|X|<\lambda$, the equation of motion is altered as
\begin{align}
  \frac{d}{dt}
  \begin{pmatrix}
    \dot{\chi} \\
    {\chi} \\
    {\varphi}
  \end{pmatrix}
=
  \begin{pmatrix}
  -\tilde \alpha   & - \delta \tilde{K}   &\tilde H_1\tilde \alpha \\
  1       &0          &0        \\
  0       &\tilde H_1    & -\tilde \alpha    \\
  \end{pmatrix}
  \begin{pmatrix}
    \dot{\chi} \\
    {\chi} \\
    {\varphi}
  \end{pmatrix}  , 
\end{align}
where 
\begin{align}
  \delta \tilde K  &= 4 \delta K S^2 \frac{a}{\lambda} \frac{2J}{\hbar^2} , 
\end{align}
and we redefined 
\begin{align}
  \chi &=  \frac{X}{\lambda} - \gamma H_0 \frac{\tilde H_1}{  \delta\tilde{K} - \tilde H_1^2 } , 
\\
   \varphi &=  \phi  + \phi_1 
  - \frac{1}{\tilde{\alpha}}  \gamma H_0 \frac{\delta\tilde{K}}{ \delta\tilde{K} - \tilde H_1^2} . 
\end{align}
 The determinant is now given by
\begin{align}
  (\varepsilon^2 + \varepsilon\tilde \alpha + \delta\tilde{K} )(\varepsilon + \tilde \alpha) 
- \tilde \alpha \tilde H_1^2  = 0 . 
\end{align}
 A positive real root exists when
\begin{align}
  \delta \tilde{K} < \tilde H_1^2 , 
\label{eq:pos_root_cond}
\end{align}
giving us the depinning condition. 
 The analytical expression of the real root of the cubic equation is given by Eq.~(\ref{eq:varepsilon}) with 
$p = \frac{1}{9} {\tilde \alpha}^2 - \frac{1}{3} \delta\tilde{K}$ and 
$q = (\frac{1}{27} {\tilde \alpha}^2 + \frac{1}{2} \tilde H_1^2 - \frac{1}{6}\delta\tilde{K}) \tilde \alpha$.

\begin{figure}[t]
  \centering
  \includegraphics[width=0.48\textwidth]{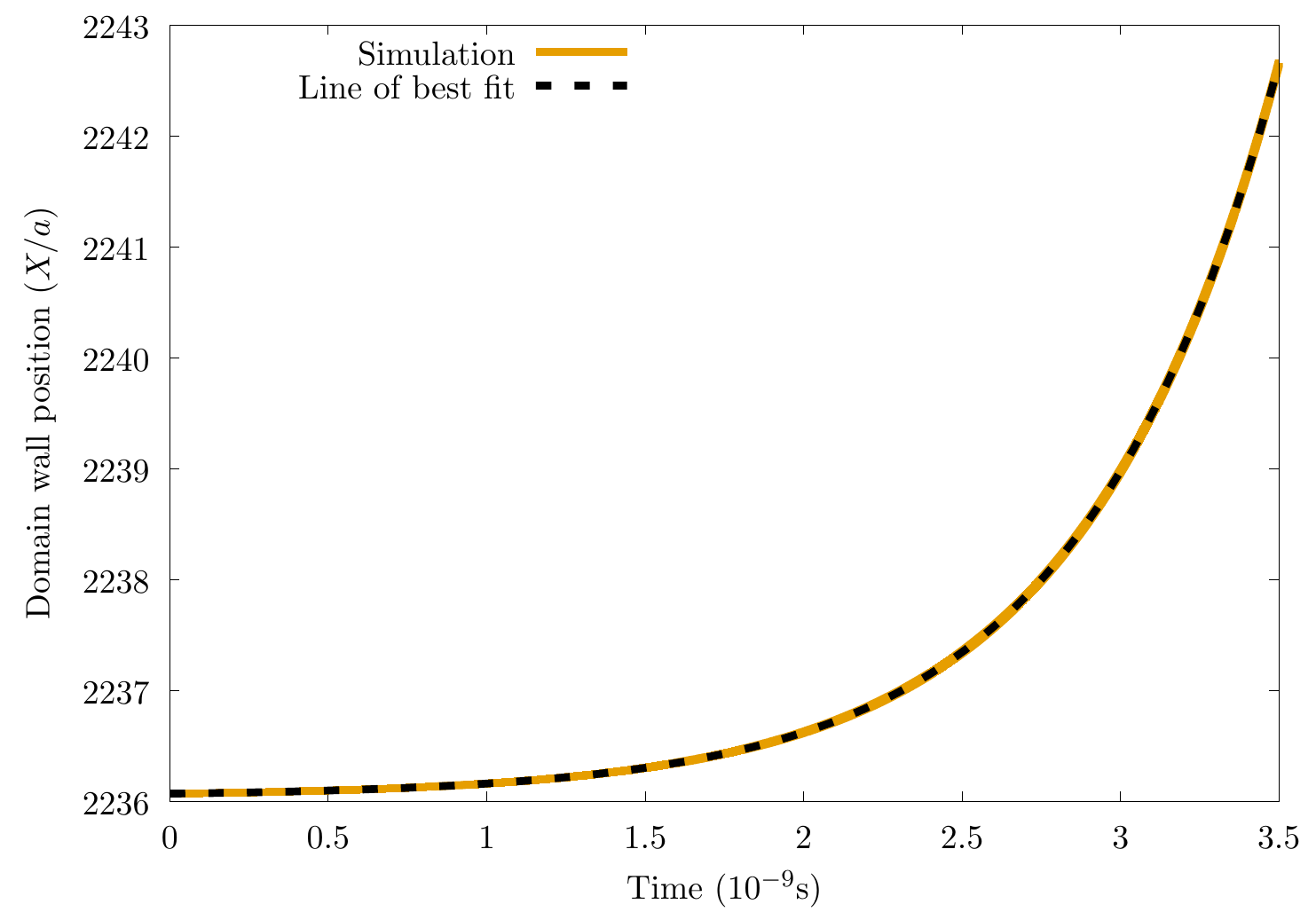}
  \caption{(Color online) 
     Domain wall position as a function of time 
under a linearly-varying magnetic field, Eq.~(\ref{eq:H(x)}). 
     The simulation result (orange line) is fitted by an exponential curve (blue dashed line), 
with exponent $\varepsilon_0 = 1.627\times10^{9}{\rm s}^{-1}$.  
    The used parameters are as follows;  
     $J =10^{3} {\rm K}$, $K = 1 {\rm K}$, $\gamma \hbar = 1{\rm K}/{\rm T}$,
      $a=10^{-10}{\rm m}$,  $H_1 = 10{\rm mT}/a$, 
      $H_0 + H_1 X(0) = H_1\lambda\times10^{-3}$ (field at the initial position $X(0)$), 
     $\alpha = 10^{-3}$, and $S=1$. 
    The system is one-dimensional and has $N=10^4$ spins (so the system size is $ L= 10^4 a = 1{\rm \mu m}$),
     and the spins at both ends are fixed upwards (${\bm S}_1 = {\bm S}_N = +\hat z$).
    The starting configuration is  Eq.~(\ref{eq:walker_dw})  
     with initial position $X(0) = 10^2 \lambda \simeq 2236 a$. 
     A time discretization of $dt = 5\times 10^{-15}{\rm s}$ is used. 
  }
\label{fig:afm_fig3}
\end{figure}

\section{Numerical Simulation}

 To test the validity of the approximations made above,  
such as the continuum description, the discarding of higher-order terms in ${\bm l}$, 
and the use of collective coordinates, 
we perform an atomistic simulation based on the equation of motion for each ${\bm S}_i$, 
\begin{align}
&   \dot{{\bm S}}_i  + \frac{\alpha}{S} {\bm S}_i \times\dot {\bm S}_i 
\nonumber \\
&\ \ \  = \hbar^{-1} \left[ J({\bm S}_{i-1}+{\bm S}_{i+1}) - 2K S_i^z \hat z  + \gamma \hbar{\bm H}_i  \right]  \times {\bm S}_i . 
\label{eq:LLG}
\end{align}
 Using the approximate domain wall solution (\ref{eq:walker_dw}) as an initial configuration, 
we solved Eq.~(\ref{eq:LLG}) under an inhomogeneous magnetic field, Eq.~(\ref{eq:H(x)}). 
 The position of the domain wall is determined by linear interpolation 
as the point at which the profile of the staggered component $(-1)^i S_i^z$ vanishes, 
and it is plotted in Fig.~\ref{fig:afm_fig3} as a function of time. 
 The values of the parameters used are described in the caption of Fig.~\ref{fig:afm_fig3}. 
 The width of the domain wall is $\lambda = a\sqrt{J/2K} \simeq 22.36 a$. 
 In accord with our analysis, the domain wall position changes exponentially with time. 
 The exponent obtained from the simulation, $\varepsilon_0 = 1.627\times10^{9}{\rm s}^{-1}$, 
is very close to the analytical result, 
$\varepsilon_0 = 1.626\times10^{9}{\rm s}^{-1}$ [Eq.~(\ref{eq:varepsilon})]. 
 The domain wall moves in the direction of stronger magnetic field.

\begin{figure}[t]
  \centering
  \includegraphics[width=0.48\textwidth]{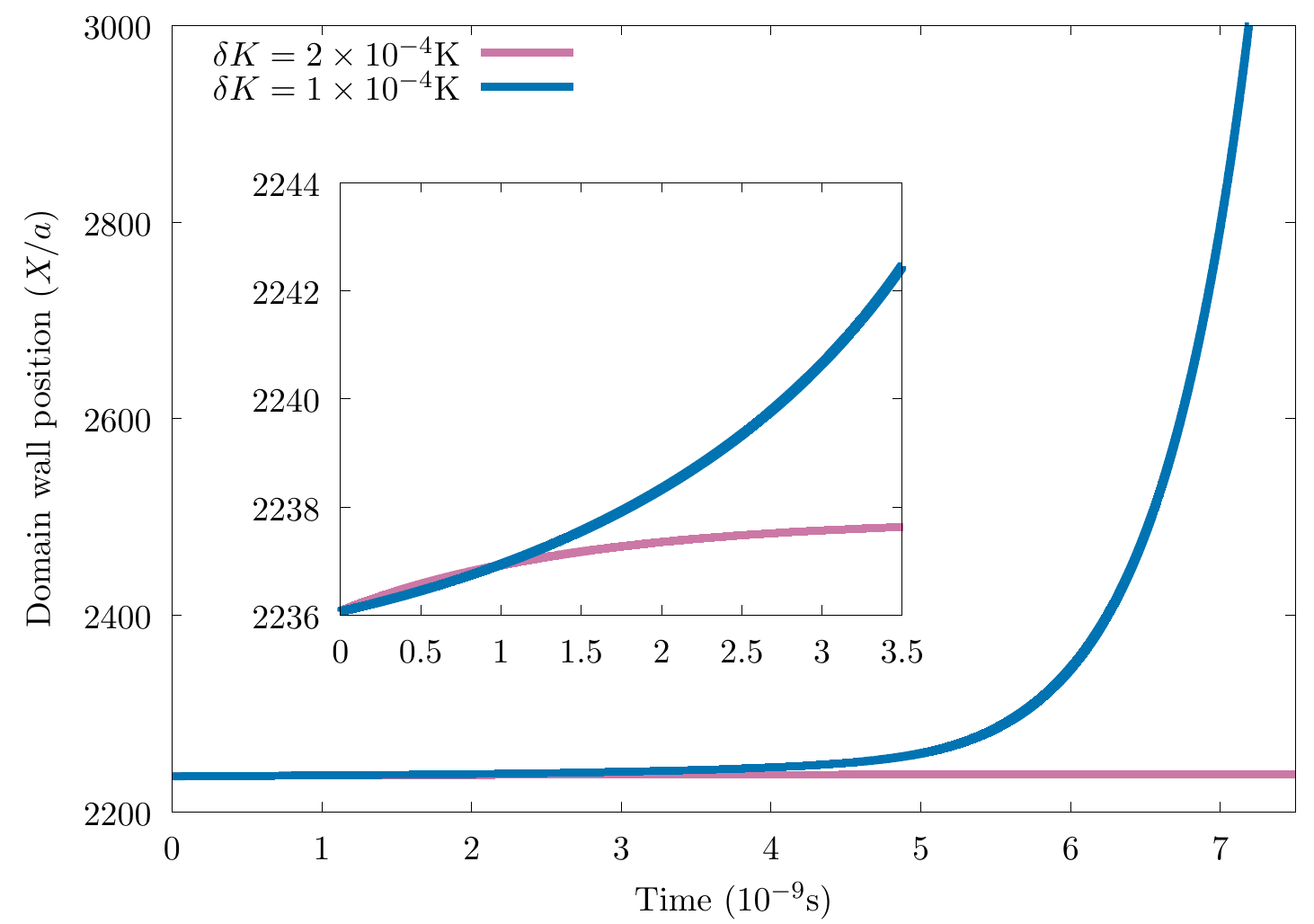}
  \caption{(Color online)
    Domain wall position as a function of time 
   in the presence of pinning potential. 
    Using the same parameters as in Fig.~\ref{fig:afm_fig3}, 
    we added a pinning potential $\delta K$ on the neighboring two sites at the initial position of the domain wall 
    (i.e. $10^2\lambda$ from the left end).
    The domain wall remains pinned for $\delta K = 2\times10^{-4}{\rm K}$ [purple (or light) line], 
    whereas it is depinned for $\delta K = 1\times10^{-4}{\rm K}$ [blue (or dark) line]. 
  The inset shows a closeup.  
  }
\label{fig:afm_fig4}
\end{figure}

 We next simulate the motion of the domain wall with pinning potential located 
at the initial position of the domain wall. 
 With the parameter values described in the caption of Fig.~\ref{fig:afm_fig3}, 
Eq.~(\ref{eq:pos_root_cond}) is satisfied when $\delta K \leq  1.398\times 10^{4}{\rm K}$.
 To test this value, we simulate the domain wall motion with 
\lq\lq strong pinning'' $\delta K = 2\times10^{-4}{\rm K}$,
and \lq\lq weak pinning'' $\delta K = 1\times10^{-4}{\rm K}$.
 As shown in Fig.~\ref{fig:afm_fig4}, 
the former pins the domain wall, while the latter cannot stop the exponential increase 
of the domain wall position.

\section{Physical Picture}

 Here, we discuss the physical mechanism of the domain wall 
actuation by an inhomogeneous magnetic field.

 First, consider a static solution. 
 Then, Eqs.~(\ref{eq:eom_X}), (\ref{eq:eom_ell_phi}), and (\ref{eq:eom_ell_theta}) 
lead to $l_\theta = \pm a/2\lambda $, $l_\phi=0$, and $X=-H_0/H_1$. 
 The first two relations show that only the artifactual texture-induced uniform moment is present, 
while the third relation tells us that the domain wall must be positioned where the magnetic field vanishes. 
 When the domain wall is placed in a finite magnetic field, the N\'eel vector starts to precess 
($\dot\phi\neq 0$) according to Eq.~(\ref{eq:eom_ell_phi}), 
 and the uniform moment $l_\theta$ develops through damping [see Eq.~(\ref{eq:eom_phi})].
 As seen from Eqs.~(\ref{eq:eom_X}) and (\ref{eq:eom_ell_theta}), 
the development of $l_\theta$ in conjunction with 
the field gradient applies a force on the domain wall;  
  like in the Stern-Gerlach experiment, the domain wall feels a force 
towards the direction with stronger magnetic field to gain Zeeman energy.
 Thus, the antiferromagnetic domain wall can be thought of as a paramagnetic particle under an 
(inhomogeneous) applied magnetic field.
 Without damping, the uniform moment is not induced by the field, 
hence there is no actuation of domain wall motion. 
 The present mechanism involves a dissipative process, hence is  
different from the purely reactive mechanisms discussed in Refs.~\cite{Tveten2016,Yuan2018}.

\vskip 3mm

\section{Reformulation}

 We have seen that the Zeeman coupling of the intrinsic magnetization 
($\sim {\bm H} \!\cdot\! \partial_x {\bm n}$) 
is nullified by the coupling of the N\'eel vector to the field gradient 
($\sim {\bm n} \!\cdot\! \partial_x {\bm H}$ in Eq.~\eqref{eq:cont_hamiltonian}).
 The reasoning behind this was given intuitively through Fig.~\ref{fig:afm_fig1}, 
which indicates that the intrinsic magnetization is not a physical quantity.

 In this section, we reformulate the theory in terms of \lq\lq physical magnetization'', 
 eliminating the intrinsic magnetization. 
 We first identify the physical magnetization by reexamining the interaction with external magnetic field, 
and therewith express the Lagrangian (Sec.~6.1). 
 The result is applied to the collective coordinates of a domain wall (Sec.~6.2).   
 Finally, the procedure is extended to general lattices (Sec.~6.3).

\subsection{Physical Magnetization}

 To identify the physical magnetization, we look at the interaction with the external 
magnetic field [the second line of Eq.~(\ref{eq:H})], and rewrite it as
\begin{align}
 2 \gamma \hbar S  \int \frac{dx}{2a} \left\{ 
  \Big( {\bm l} + \frac{a}{2} \partial_x{\bm n} \Big) \cdot {\bm H}
        -   \frac{a}{2} \, \partial_x ({\bm H} \cdot {\bm n} )  \right\} . 
\label{eq:Zeeman}
\end{align}
 The first term is the Zeeman coupling in the bulk, and 
the second (total-derivative) term describes that at the edges. 
 Therefore, the physical magnetization is identified to be $- \gamma s_0 \tilde{\bm l}$, with 
\begin{align}
  \tilde {\bm l} &\equiv  {\bm l} + \frac{a}{2}(\partial_x{\bm n})  . 
\label{eq:ell_physical} 
\end{align}
 This is the uniform moment with the intrinsic magnetization subtracted, 
and agrees with Haldane's definition\cite{Haldane1983} 
(according to the analysis made in Ref. \cite{Tveten2016}). 
 In terms of $\tilde {\bm l}$, the Lagrangian density, Eq.~(\ref{eq:L_density}), is simplified as 
\begin{align}
  {\cal L}  &=  s_0 \tilde {\bm l} \cdot ( {\bm n} \times \dot {\bm n} - \gamma   {\bm H} ) 
\nonumber \\
&\ \  \   - s_0  \frac{2JS}{\hbar}  
  \Bigg\{  \tilde {\bm l}^{\, 2} + \frac{a^2}{4} (\partial_x{\bm n})^2
      - \frac{K}{2J} (n^z)^2
  \Bigg\} . 
\end{align}
 Here, we dropped the \lq\lq topological'' term 
$\sim \partial_x {\bm n} \cdot ( {\bm n} \times \dot {\bm n})$ 
since it does not affect the equation of motion. 
 Note that the exchange stiffness constant of the N\'eel vector (the coefficient of $(\partial_x{\bm n})^2$)
has been reproduced correctly (without eliminating the uniform moment), 
and the \lq\lq sublattice symmetry'' $(\tilde {\bm l}, {\bm n}) \to (\tilde {\bm l}, -{\bm n})$ 
has been recovered. 
 Note also that the constraints are preserved, 
$\tilde {\bm l} \cdot {\bm n} = 0$ and ${\bm n}^2+{\tilde {\bm l}}^2 = 1$, within the exchange approximation. 
 These suggest that the theory is simplified if reformulated in terms of $\tilde {\bm l}$.

 To complete this program, we need to examine the damping term. 
 If the dissipation function has the form, 
\begin{align}
  W &= s_0 \int  dx 
 \left(  \frac12 \alpha_l \dot {\tilde {\bm l}}^2 + \frac12 \alpha_n \dot {\bm n}^2  \right) , 
\label{eq:rayleigh_dissipation_neel_correct}
\end{align}
in terms of $\tilde {\bm l}$ 
[instead of ${\bm l}$ as in Eq.~(\ref{eq:rayleigh_dissipation_neel})], 
the damping term is also simplified. 
 To show that this is indeed the case, it is sufficient to observe that the spins ${\bm S}_i$ couple 
to other degrees of freedom (\lq\lq environment'') through $(\tilde {\bm l}, {\bm n})$ 
rather than $({\bm l}, {\bm n})$. 
 As an example, let us consider the s-d exchange coupling to conduction electrons,  
\begin{align} 
 H_{\rm sd} &= - J_{\rm sd} \sum_i {\bm S}_i \cdot {\bm \sigma}_i  , 
\end{align}
where ${\bm \sigma}_i$ is the electron spin at site $i$, and $J_{\rm sd} $ is the coupling constant. 
 This has the same form as the Zeeman coupling (${\bm H}_i \to {\bm \sigma}_i$), 
and we can proceed in exactly the same way as Eq.~(\ref{eq:Zeeman}). 
 By noting that ${\bm \sigma}_i$ may have staggered component as well, 
we obtain the s-d coupling in the continuum approximation as 
\begin{align} 
 H_{\rm sd} 
&= - J_{\rm sd} S \int d {\bm r} \, 
\left\{ 
  \Big( {\bm l} + \frac{a}{2} \partial_x{\bm n} \Big) 
\cdot {\bm \sigma}_l 
         + {\bm n} \cdot  {\bm \sigma}_n  \right\} , 
\end{align}
where ${\bm \sigma}_l$ and ${\bm \sigma}_n$ are the uniform and staggered components 
of the electron spin density. 
 (For simplicity, we dropped the total derivative terms.) 
 As seen, there is a \lq\lq correction'' $a \, \partial_x {\bm n}/2$ in the first term, 
and the coupling to the electrons occurs through $(\tilde {\bm l}, {\bm n})$, 
instead of $({\bm l}, {\bm n})$. 
 (Precisely speaking, the \lq\lq corrections'' arise symmetrically between ${\bm l}$ and ${\bm n}$, 
but in the second term, we adopted the exchange approximation, 
  ${\bm n} + \,  a \, \partial_x {\bm l}/2 \simeq {\bm n}$.) 
 Therefore, the resulting Gilbert damping, or the dissipation function, should have the form of 
Eq.~(\ref{eq:rayleigh_dissipation_neel_correct}) through $(\tilde {\bm l}, {\bm n})$.

 Recently, we have conducted explicit calculations of Gilbert damping and spin-transfer torques, 
and found that the expectation value of ${\bm\sigma}_n$ is odd in $\bm n$ 
while that of ${\bm\sigma}_l$ is even.\cite{Nakane_current}  
 Thus, the spin torques resulting from the s-d exchange interaction also possess 
the sublattice symmetry under $(\tilde{\bm l},{\bm n})\to(\tilde{\bm l},-{\bm n})$.

\subsection{Domain Wall}

 For a domain wall in the collective coordinate description, 
the physical magnetization, Eq.~(\ref{eq:ell_physical}), is given by  
\begin{align}
  \tilde l_\theta &\equiv  l_\theta \mp \frac{a}{2\lambda}  ,
\end{align}
and $l_\phi $ (not altered). 
 With these, the equations of motion are written as 
\begin{align}
 \pm \dot{l}_\phi &=  - \alpha_n \dot{X} / \lambda + \gamma H_1 \lambda \tilde  l_\theta   , 
\label{eq:eom_X_final}
\\
   \dot{ \tilde l}_\theta  &=    \alpha_n \dot{\phi}  , 
\label{eq:eom_phi_final}
\\
   \dot \phi  &=  - (4J S /\hbar) \, \tilde l_\theta + \gamma (H_0 + H_1 X) , 
\label{eq:eom_ell_phi_final}
\\
     \dot X &= \pm (4J S /\hbar)  \lambda l_\phi . 
\label{eq:eom_ell_theta_final}
\end{align} 
 The sign $\pm$ represents the topological charge. 
 We see that a static domain wall has 
$\tilde l_\theta = \gamma \hbar H_0 /4JS$ for $H_1 = 0$, 
and $\tilde l_\theta = 0$ for $H_1 \ne 0$.

\subsection{General Case}

 The procedure described in Sec.~6.1 can be generalized to arbitrary bipartite lattices. 
 We assume a nearest-neighbor exchange interaction $J$ and a uniaxial magnetic anisotropy $K$. 
 Taking the unit cell along the $x$ direction, 
the Hamiltonian density (without Zeeman coupling terms) is calculated as 
\begin{align}
  {\cal H}_{D,\nu}  &=  s_0  \frac{\nu JS}{\hbar}  
  \Bigg\{  \tilde {\bm l}^{\, 2} + \frac{a^2}{4D} \sum_{i=1}^D (\partial_i {\bm n})^2
      - \frac{K}{\nu J} (n^z)^2
  \Bigg\} , 
\end{align}
where $\tilde {\bm l}$ is given by Eq.~(\ref{eq:ell_physical}), 
$D$ is the space dimensionality, and $\nu$ is the number of nearest-neighbor sites. 
 For example, $(D, \nu) = (2,3), (2,4)$, and (3,6) for honeycomb, square, and simple cubic lattices, respectively. 
 The Zeeman coupling is the same as Eq.~(\ref{eq:Zeeman}), 
hence $\tilde {\bm l}$ is identified as the physical magnetization. 
 The Lagrangian density is given by 
${\cal L} = s_0 \tilde {\bm l} \cdot ( {\bm n} \times \dot {\bm n} - \gamma   {\bm H} ) - {\cal H}_{D,\nu}$.

\section{Summary}

 We have investigated the motion of an antiferromagnetic domain wall under inhomogeneous magnetic field. 
 Starting from the lattice Heisenberg model with antiferromagnetic exchange coupling, 
easy-axis anisotropy, and Zeeman coupling, we constructed a continuum model 
by closely following Ref.~\cite{Tveten2016}. 
 We first retrieved a term that was missing in Ref.~\cite{Tveten2016}, 
 in which the N\'eel vector couples to the field gradient.  
 We have shown that this retrieved term nullifies the previously demonstrated coupling 
of the intrinsic magnetization, attributed to the N\'eel spin texture, to the magnetic field, 
and found an alternative mechanism for domain wall motion actuated by an inhomogeneous field.

 As a supplemetary discussion, we pointed out that the uniform moment ${\bm l}$ 
defined by Eq.~(\ref{eq:nl}) contains unphysical component (intrinsic magnetization). 
 We have reformulated the theory by properly defining the physical magnetization.

\acknowledgment 
 We thank T. Funato, Y. Imai, K. Nakazawa, T. Yamaguchi, A. Yamakage, K. Yamamoto, and Y. Yamazaki 
for helpful discussion. 
 This work is supported by JSPS KAKENHI Grant Numbers JP15H05702, JP17H02929 and JP19K03744. 
 JJN is supported by a Program for Leading Graduate Schools ``Integrative Graduate Education and Research in Green Natural Sciences'' 
 and Grant-in-Aid for JSPS Research Fellow Grant Number 19J23587.

\end{document}